\documentclass[twoside,twocolumn,9pt]{article}
\usepackage{extsizes}
\usepackage[super,sort&compress,comma]{natbib} 
\usepackage[version=3]{mhchem}
\usepackage[left=1.5cm, right=1.5cm, top=1.785cm, bottom=2.0cm]{geometry}
\usepackage{balance}
\usepackage{widetext}
\usepackage{multicol}
\usepackage{times,mathptmx}
\usepackage{sectsty}
\usepackage{graphicx} 
\usepackage{lastpage}
\usepackage[format=plain,justification=raggedright,singlelinecheck=false,font={stretch=1.125,small,sf},labelfont=bf,labelsep=space]{caption}
\usepackage{float}
\usepackage{fancyhdr}
\usepackage{fnpos}
\usepackage[english]{babel}
\usepackage{array}
\usepackage{droidsans}
\usepackage{charter}
\usepackage[T1]{fontenc}
\usepackage[usenames,dvipsnames]{xcolor}
\usepackage{setspace}
\usepackage[compact]{titlesec}
\usepackage{amsthm,amsmath,amssymb}
\usepackage[dvipsnames]{xcolor}

\usepackage{epstopdf}

\usepackage[title]{appendix}
\usepackage{subfigure}

\definecolor{cream}{RGB}{222,217,201}

\begin{document}

\pagestyle{fancy}
\thispagestyle{plain}
\fancypagestyle{plain}{

\renewcommand{\headrulewidth}{0pt}
}

\makeFNbottom
\makeatletter
\renewcommand\LARGE{\@setfontsize\LARGE{15pt}{17}}
\renewcommand\Large{\@setfontsize\Large{12pt}{14}}
\renewcommand\large{\@setfontsize\large{10pt}{12}}
\renewcommand\footnotesize{\@setfontsize\footnotesize{7pt}{10}}
\makeatother

\renewcommand{\thefootnote}{\fnsymbol{footnote}}
\renewcommand\footnoterule{\vspace*{1pt}%
\color{cream}\hrule width 3.5in height 0.4pt \color{black}\vspace*{5pt}} 
\setcounter{secnumdepth}{5}

\makeatletter 
\renewcommand\@biblabel[1]{#1}            
\renewcommand\@makefntext[1]%
{\noindent\makebox[0pt][r]{\@thefnmark\,}#1}
\makeatother 
\renewcommand{\figurename}{\small{Fig.}~}
\sectionfont{\sffamily\Large}
\subsectionfont{\normalsize}
\subsubsectionfont{\bf}
\setstretch{1.125} 
\setlength{\skip\footins}{0.8cm}
\setlength{\footnotesep}{0.25cm}
\setlength{\jot}{10pt}
\titlespacing*{\section}{0pt}{4pt}{4pt}
\titlespacing*{\subsection}{0pt}{15pt}{1pt}

\fancyfoot{}
\fancyfoot[RO]{\footnotesize{\sffamily{1--\pageref{LastPage} ~\textbar  \hspace{2pt}\thepage}}}
\fancyfoot[LE]{\footnotesize{\sffamily{\thepage~\textbar\hspace{3.45cm} 1--\pageref{LastPage}}}}
\fancyhead{}
\renewcommand{\headrulewidth}{0pt} 
\renewcommand{\footrulewidth}{0pt}
\setlength{\arrayrulewidth}{1pt}
\setlength{\columnsep}{6.5mm}
\setlength\bibsep{1pt}

\makeatletter 
\newlength{\figrulesep} 
\setlength{\figrulesep}{0.5\textfloatsep} 

\newcommand{\topfigrule}{\vspace*{-1pt}%
\noindent{\color{cream}\rule[-\figrulesep]{\columnwidth}{1.5pt}} }

\newcommand{\botfigrule}{\vspace*{-2pt}%
\noindent{\color{cream}\rule[\figrulesep]{\columnwidth}{1.5pt}} }

\newcommand{\dblfigrule}{\vspace*{-1pt}%
\noindent{\color{cream}\rule[-\figrulesep]{\textwidth}{1.5pt}} }

\newcommand{\hC}{\hat{C}}
\newcommand{\hm}{\hat{m}}

\makeatother

\twocolumn[
  \begin{@twocolumnfalse}
\vspace{3cm}
\sffamily
\begin{tabular}{m{4.5cm} p{13.5cm}}

 & \noindent\LARGE{\textbf{Growth of a bubble cloud in CO$_2$-saturated water in microgravity$^\dag$}} \\
\vspace{0.3cm} & \vspace{0.3cm} \\

 & \noindent\large{Patricia Vega-Mart\'{\i}nez\textit{$^{a}$}, Javier Rodr\'{\i}guez-Rodr\'{\i}guez,$^{\ast}$\textit{$^{a}$} and Devaraj van der Meer\textit{$^b$}} \\

 & \noindent\normalsize{The diffusion-driven growth of a dense cloud of bubbles immersed in a gas-supersaturated liquid is a problem that finds applications in several modern technologies such as solvent-exchange micro-reactors, nanotechnology or the manufacturing of foamy materials. However, under Earth's gravity conditions, these dynamics can only be observed for a very limited time if the cloud is not attached to a surface, due to the action of buoyancy, i.e. of gravity effects. Here, we present experimental observations of the time evolution of dense bubble clouds growing in CO$_2$-supersaturated water in microgravity conditions. We report the existence of three regimes where the bubble cloud exhibits different growth rates. At short times, each bubble grows independently following the Epstein--Plesset equation. Later on, bubbles start to interact with each other and their growth rate diminishes as they compete for the available CO$_2$. When this happens, the growth rate slows down. This occurs earlier the deeper the bubble is in the cloud. Finally, at long times, only those bubbles on the husk continue growing. These regimes may be qualitatively described by a mathematical model where each individual bubble grows in the presence of a constellation of point mass sinks. Despite the model being only valid for dilute bubble clouds, its predictions are consistent with the experimental observations, even though the bubble clouds we observe are rather dense.}\\

\end{tabular}

\end{@twocolumnfalse} \vspace{0.6cm} ]


\renewcommand*\rmdefault{bch}\normalfont\upshape
\rmfamily
\section*{}
\vspace{-1cm}


\footnotetext{\textit{$^{a}$~Fluid Mechanics Group, Carlos III University of Madrid, Legan\'es, Spain. E-mail: bubbles@ing.uc3m.es}}
\footnotetext{\textit{$^{b}$~Physics of Fluids group. University of Twente, Enschede, The Nederlands. }}

\footnotetext{\dag~Electronic Supplementary Information (ESI) available: [details of any supplementary information available should be included here]. See DOI: 10.1039/b000000x/}


\section{Introduction}

The diffusion-driven dynamics of a dense cloud of bubbles (or drops) have received lately a renewed attention due to their applications in modern chemical processes such as solvent extraction \cite{Peng_etalJCIS2018} and in nanoscience \cite{Zhu_etalSoftMatter2018}. Besides these immediate applications, understanding these diffusive dynamics also sheds light on certain processes taking place in foams, such as coarsening or coalescence. Indeed, although many other physical effects are involved in the evolution of a wet foam, still diffusion-related phenomena play a central role\cite{Langevin2017}. However, despite the diffusive interaction between nearby bubbles in bubbly liquids or foams having been studied for a long time, still there is a need for quantitative predictive models\cite{Michelin_etalPRF2018}. In part, this is a consequence of the difficulties that arise in isolating purely diffusive effects in experiments. In fact, the large difference in density between the bubbles and the liquid makes gravity effects manifest rather early in the evolution of foams or bubble clouds\cite{Langevin2017}. 

In foams it is hard to disentangle the role of liquid drainage (a gravity-related effect) from other effects associated to the transport of gas\cite{garcia_moreno_etalSofMatter2011} or anti-foaming agents\cite{yazhgur2015antifoams} in the rate of bubble coalescence. Moreover, drainage turns wet foams into dry ones very fast, so the former are hard to characterize experimentally\cite{Langevin2017}. Even in a situation where diffusion is the leading effect, such as the growth of a bubble cloud in a gas-supersaturated liquid, gravity effects become of comparable importance very quickly. For example, a free cloud of sub-millimetric bubbles exhibits a purely diffusion-driven growth only for a very limited time (hundreds of miliseconds), even for high gas concentrations such as those found for instance in beer or champagne, as shown by \citet{rodriguez2014}. In this study, a dense bubble cloud was produced in the bulk of a CO$_2$-saturated liquid (beer) and its evolution studied using high-speed imaging. This evolution could be divided into three stages: in a first one, which lasts only about 10 ms, the size of the bubbles increases with the square root of time, as predicted by the Epstein-Plesset equation for the case of a single isolated bubble\cite{EpsteinPlessetJCP1950}. Thus, it is reasonable to assume that in this stage bubbles grow without interacting much with each other. Then, another stage follows where bubbles moderate their growth, due to the depletion of the dissolved gas in the space between bubbles. These two stages, where the dynamics are driven by diffusive bubble growth, come to an end when gravity effects become important and make the bubble cloud rise, which occurs as early as about 100 ms even for a bubble cloud as small as about 1 mm in diameter. Of course, bubbles could be prevented from rising by keeping them pinned to a substrate. However, the pinning introduces additional complexities that obscure, to some extent, diffusive effects. Moreover, in some situations, gravity plays a role even if the bubbles remain pinned to a substrate. In the case of sessile bubbles in CO$_2$-supersaturated water, \citet{Enriquez_etalJFM2014} and \citet{Moreno_soto_etalJFM2019} have shown that, although bubbles are not allowed to move, the depletion of CO$_2$ that takes place around them induces density gradients in the liquid that ultimately trigger a convection plume, which in turn dominates the flow and overcomes the effect of diffusion.

Therefore, avoiding or minimizing gravity effects is essential to study the purely diffusion-driven dynamics of wet foams or bubble clouds. Moreover, besides this purely fundamental interest, there are situations in which bubble clouds or wet foams evolve in microgravity or, at least, in reduced gravity conditions. For example, in the field of planetary physics, the diffusive growth of bubbles is responsible for the loss of Helium in meteorites 
\cite{stuart1999}. Thus, modeling the evolution of Helium content in these objects is important to understand the history of the formation of our solar system. In the Moon, rocks with a highly vesicular structure, such as the "Seatbelt Rock", have been found (see figure 4 in \citet{JolliffRobinsonPT2019}). This structure suggests that the rock has formed by solidification of a dense cloud of bubbles containing volatile gases, as a result of ancient volcanism. From a more technological point of view, the behavior of foams in reduced gravity conditions is of interest for the development of advanced manufacturing processes in space, as pointed out by \citet{Koursari_etalMGST2019}. These authors study experimentally the drying of a wet foam in the absence of gravity. Since on Earth gravity is essential to drain out the water filling the interstices between bubbles, in space they use a porous medium to soak up the liquid. Note that, if bubbles were allowed to grow by supersaturating the liquid with a gas, their growth could push out the interstitial water without the need of an external absorbing body.

In summary, studying experimentally the diffusive interaction of free growing bubbles in dense bubble clouds or in wet foams at long times requires avoiding the effect of gravity. This has been done in several studies in the case of foams\cite{garcia_moreno_etalSofMatter2011, garcia_moreno_etalSoftMatter2014,garcia_moreno_NatComm2019, Koursari_etalMGST2019} but, to the best of our knowledge, there are no studies focusing on dense bubble clouds growing diffusively in gas-supersaturated liquids in microgravity. With this motivation in mind, we have carried out experiments in which dense bubble clouds grow in CO$_2$-supersaturated water in microgravity conditions. These experiments were performed in the drop tower of the German Center of Applied Space Technology and Microgravity (ZARM) using an improved version of the facility described by \citet{VegaMartinez_etalMGST2017}. Besides, inspired by these experiments, we have put together a mathematical model of the growth of a bubble cloud in a supersaturated liquid that is able to explain, albeit in a qualitative fashion, the features of the time evolution of bubble clouds observed in the experiments.

\section{\label{sec:experiments}Experimental setup and procedure}

\begin{figure}[h]
    \centering
        \includegraphics[width=\columnwidth]{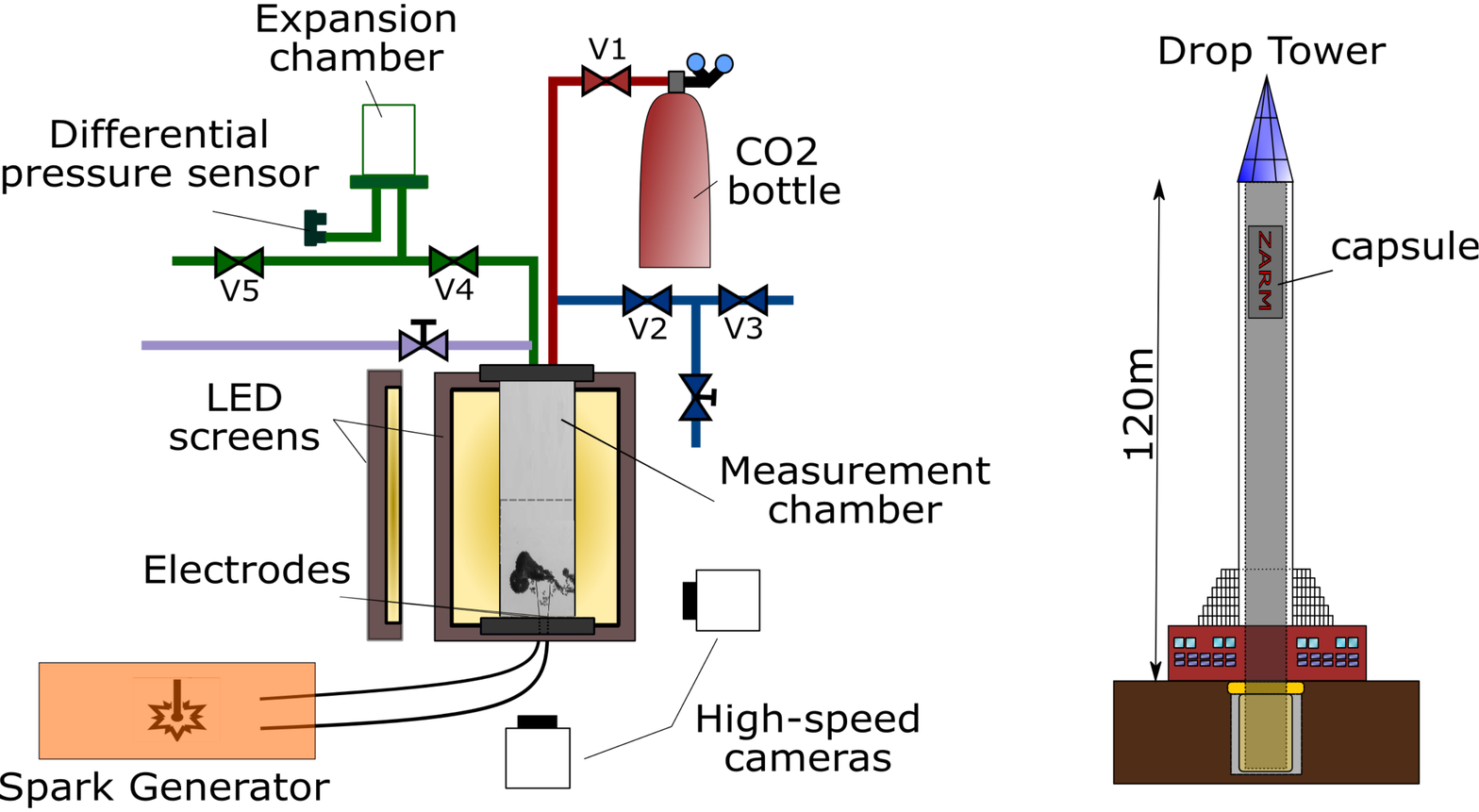}
    \caption{Layout of the experimental setup and the drop tower facility. The two high-speed cameras form an angle of 90$^{\,\rm o}$. The chamber has a cylindrical shape, 24.4 mm of diameter and 200 mm of height. The bubble cloud is generated at the center of the chamber to avoid the effect of walls. The right panel is a sketch of the capsule inside the drop tower (not to scale)}
    \label{fig:setup}
\end{figure}

The experiment aims at generating a dense bubble cloud in a supersaturated liquid, in our case CO$_2$-saturated water, and then observe its evolution using high-speed imaging in microgravity conditions. Figure~\ref{fig:setup} shows the sketch of the experimental setup, which is based on that described in \citet{VegaMartinez_etalMGST2017}. This experiment is fitted into a capsule that is then dropped inside the drop tower of the German Center of Applied Space Technology and Microgravity (ZARM). This tower has total height of 146 meters and contains in its interior a 120-m-tube where the capsule falls in near vacuum for about 4 s. This generates a microgravity environment for the duration of the drop. Then the capsule decelerates and comes to a stop in a pool filled with polystyrene pellets.

The measurement chamber, where the bubble cloud will form and grow, is filled with carbonated water. This chamber is immersed in a rectangular tank filled with degassed water to minimize optical distortion. The top wall of the chamber is connected to a pressurization and depressurization system. 
At the bottom of the measurement chamber there are two electrodes connected to the spark generator. The electrodes are two thin (100 $\mu$m) copper wires that touch each other near their tips. The spark generator is a discharge circuit similar to that described in \citet{Willert2010} but replacing the LEDs by the electrodes, as suggested by \citet{Goh2013}. In our experiments, the discharge has a peak between 50-100 A and lasts for about 10 $\mu$s. The discharge induces cavitation, and the collapse of the imploding cavitation bubble generates the bubble cloud, the growth of which is the target of the experiment. To measure the time evolution of the shape and size of the cloud, two high-speed cameras, suitable for use in the drop tower's capsule, image the bubble cloud from two perpendicular directions at 2000 fps. They produce images of 512$\times$512 pixels and a spatial resolution of 25 $\mu$m/pixel. As can be seen in the appendix, the information provided by the two cameras is qualitatively similar for all drops. For that reason, in the discussion of the experimental results we will only use images taken from camera 1.

The supersaturated water is prepared in a pressurized facility built to make carbonated water on-the-site which is subsequently transferred to the measurement chamber at a pressure higher than the saturation one, such as no gas is lost during the process. This procedure is similar to that followed by \citet{Enriquez2013}.

\begin{figure*}[t]
\centering
\includegraphics[width=\textwidth]{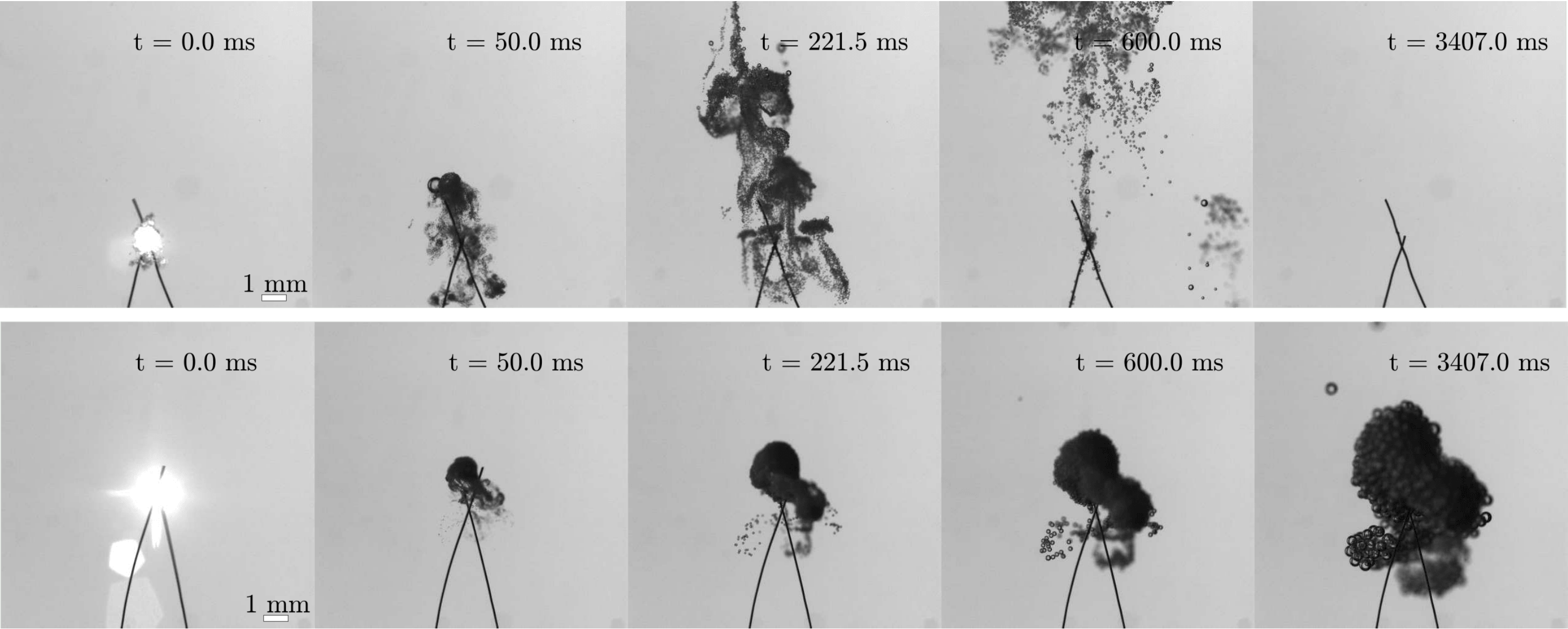}
\caption{Comparison of experiments where the bubble cloud develops with (upper row) and without gravity (lower row). See movie in supplemental material.}
\label{fig:panel}
\end{figure*}

Figure~\ref{fig:panel} compares the evolution of a bubble cloud in microgravity ($10^{-6}$ $g$) and 1 $g$ conditions from its generation (the spark) to the end of the experiment, approximately 3.4 seconds later. During the first 100 ms, both tests show a similar diffusion-driven growth. Subsequently, in the 1 $g$ test, the effects of the gravity become abundantly clear: the cloud starts a buoyancy-induced rising motion as can be seen in the first row of the figure. In the microgravity test, however the bubble cloud continues growing by diffusion and no net motion is observed. For the model developed in the next section it is important to point out that, despite the high bubble density of the cloud, we observe that very few bubbles coalesce during the experiments. Thus, in a first approximation the number of bubbles may be regarded as constant.

\section{Mathematical model of transient mass transfer in a bubble cloud}

\begin{figure}[h!]
    \centering
    \includegraphics[width=0.9\columnwidth]{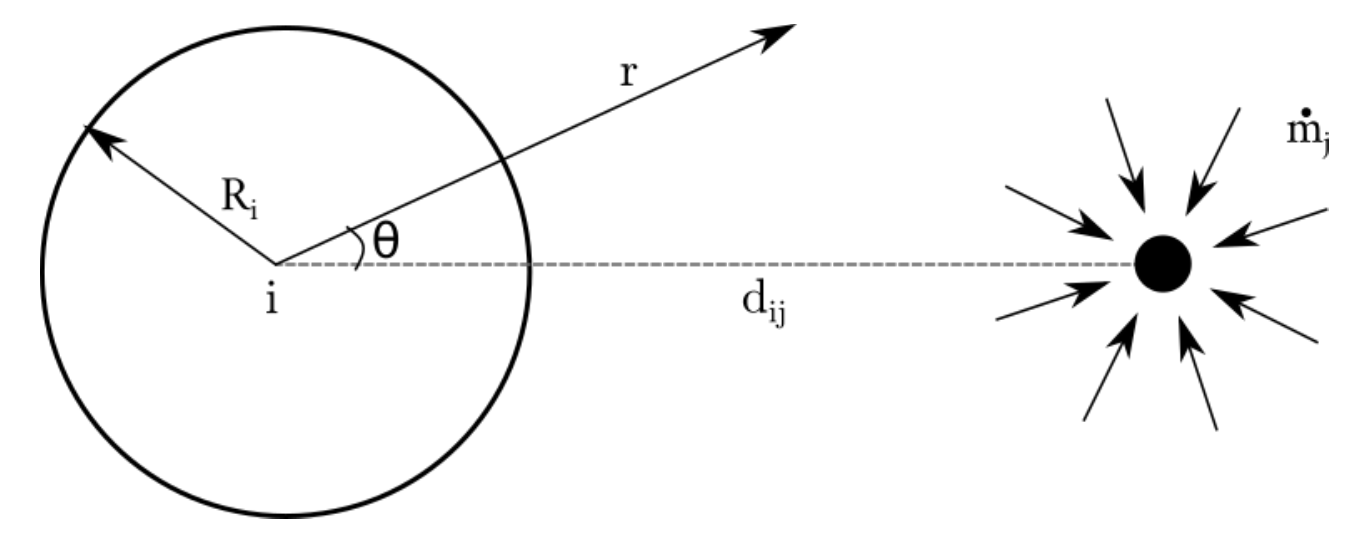}
    \caption{\label{fig:bubble_and_sink}Bubble of radius $R_i$ in the presence of a mass sink of intensity $\dot{m}_j$ located a distance $d_{ij}$ apart.}
\end{figure}

Let us consider a spherical bubble of radius $R_i$ surrounded by a cloud of other bubbles of radii $R_j$, with $j = 1..N$ and $j \ne i$. We introduce here three hypotheses. First, from the point of view of mass transfer, one bubble sees the others in the cloud as mass sinks of intensity $\dot{m}_j$. Second, we do not consider surface tension effects. Third, advective effects caused by bubble growth are neglected in the gas transport equation. This means that the concentration of dissolved gas obeys the non-convective heat equation. Due to the linear nature of this equation, the interaction of the $i$-th bubble with the surrounding ones can be treated by superposition. For this reason, we consider the growth of a bubble, labelled $i$, in presence of a point mass sink of intensity $\dot{m}_j$ a distance $d_{ij}$ apart from its center (see figure \ref{fig:bubble_and_sink}). The problem we will solve now is to determine the gas mass flux entering bubble $i$ in these conditions. This problem exhibits cylindrical symmetry around the line connecting the centers of both bubbles, so the concentration field obeys
\begin{equation}
    \partial_t C - D\left[\frac{1}{r^2}\partial^2_{rr}\left(r^2 \, C\right) + \frac{1}{r^2\sin\theta}\partial_\theta\left(\sin\theta\partial_\theta C\right)\right] = -\dot{m}_j\delta\left[\vec{x}-d_{ij}\vec{e}_x\right].
    \label{eq:time_eqnC}
\end{equation}
This equation must be subject to the boundary conditions at the bubble surface, $r=R_i$, and in the far field $r \rightarrow \infty$. At the bubble surface the concentration is given by Henry's law, in other words $C(R, t) = C_s = k_H P_s$, where $k_H$ is Henry's constant and $P_s$ the ambient pressure,  whereas in the far field $C(\vec{x}, t) = C_\infty$. Note that, although equation (\ref{eq:time_eqnC}) is linear, the fact that the radius of the $i$-th bubble, $R_i$, depends on time effectively couples the computation of the mass fluxes of the different bubbles, as every bubble $j$ affects the concentration field induced by the others through the boundary condition at $r = R_i$. Nonetheless, as will be explained below, this effect can be neglected with the assumptions adopted here.

The solution of the problem will be sought by exploiting its linear nature to decompose it into a spherically-symmetric one that satisfies these boundary conditions plus another one that accounts for the effect of the point sink but that has homogeneous boundary conditions. The former solution recovers the mass flux given by the Epstein--Plesset equation\cite{EpsteinPlessetJCP1950}, $\dot{m}_{i0}$. To calculate the latter, denoted by $\dot{m}_{ij}$, we transform equation (\ref{eq:time_eqnC}) into the Laplace plane. In what folows, $\hC(r, s)$ and $\hm_{j}(s)$ are the Laplace transforms of $C(r, t)$ and $\dot{m}_{j}(t)$, respectively. With this notation, equation (\ref{eq:time_eqnC}) turns into
\begin{equation}
    s\,\hC - D\left[\frac{1}{r^2}\partial^2_{rr}\left(r^2 \, \hC\right) + \frac{1}{r^2\sin\theta}\partial_\theta\left(\sin\theta\partial_\theta \hC\right)\right] = -\hm_{j}\delta\left[\vec{x}-d_{ij}\vec{e}_x\right].
    \label{eq:laplace_eqnC}
\end{equation}
To transform the time derivative of the concentration field we take into account that, with the decomposition explained above, the concentration field induced by an individual mass point sink has homogeneous initial conditions, thus $\partial C/\partial t$ transforms to $s\,\hat{C}$.

The general solution to this equation is\cite{WuSF2015}:
\begin{equation}
\begin{split}
    \hC & = -\frac{\hm_{j}}{4\pi D}\frac{\exp\left(-\sqrt{s/D}\sqrt{d_{ij}^2 + r^2 - 2d_{ij}r\cos\theta}\right)}{\sqrt{d_{ij}^2 + r^2 - 2d_{ij}r\cos\theta}} +\\
    & + \sum_{l=0}^\infty\left[a_l\,h_l^{(1)}(\sqrt{-s/D}\,r) + b_l\,h_l^{(2)}(\sqrt{-s/D}\,r)\right]\,P_l(\cos\theta),
\end{split}
\label{eq:hC_general_solution}
\end{equation}
where $h_l^{(1)}$ and $h_l^{(2)}$ are Hankel's functions of first and second kind, respectively. Note that, since they contain exponential functions of imaginary exponent, the minus sign inside the square root of their arguments guarantees that they take real values. Although to compute the concentration field it is necessary to determine the full set of coefficients $(a_l, b_l)$, the calculation of the mass flux $\hm_{ij}$ that sink $j$ induces on the bubble only requires knowledge of the coefficients $a_0$ and $b_0$. Indeed, the mass transfer due to higher-order harmonics is zero, as
\begin{equation}
    \dot{m}_{ij} = -2\pi R_i^2 D \, \int_0^\pi \left.\frac{\partial C}{\partial r}\right|_{r=R}\,\sin\theta\,\mathrm{d}\theta \sim \int_0^\pi P_l(\cos\theta)\,\sin\theta\,\mathrm{d}\theta,
    \label{eq:mass_flux}
\end{equation}
with the last expression being zero for $l \ge 1$. Moreover, the coefficient $b_0$ must vanish, as the function $h_0^{(2)}\left(\sqrt{-s/D}r\right)$ is unbounded as $r \rightarrow \infty$.

To obtain $a_0$ we introduce (\ref{eq:hC_general_solution}) into (\ref{eq:laplace_eqnC}) and project onto $P_0(\cos\theta)$ to find
\begin{equation}
\begin{split}
    0 = & -\frac{\hm_j}{4\pi D}\,\int_{-1}^1{ \frac{\exp\left(-\sqrt{\displaystyle \frac{s}{D}}\sqrt{d_{ij}^2 + R_i^2 - 2d_{ij}R_i x}\right)}{\sqrt{d_{ij}^2 + R_i^2 - 2d_{ij}R_i x}}\,\mathrm{d}x} + \\
    & + 2 a_0 h_0^{(1)}\left(\sqrt{-s/D}\,R_i\right),
\end{split}
\end{equation}
where the variable change $x = \cos\theta$ has been introduced. Here it is good to note that the expression of Eq. (5) is hybrid, as it mixes expressions in Laplace space with the radius of bubble $i$, which generally is a function of time, $R_i(t)$. We will however ignore this fact and treat $R_i$ as a constant in Laplace space to make analytical progress. Physically, this amounts to saying that we assume that the time scale with which the concentration field responds to changes in the configuration is much smaller than that at which the bubble grows. This is an assumption that is quite similar to the ``frozen bubble'' assumption that was employed in solving the single bubble problem in the original paper by Epstein \& Plesset \cite{EpsteinPlessetJCP1950}.
Evaluating the integrals, and taking into account $h_0^{(1)}(\sqrt{-s/D}\,R_i) = -e^{-\sqrt{s/D}\,R_i}/\sqrt{s/D}\,R_i$,
\begin{equation}
    0 = \frac{\hm_j}{4\pi D}\frac{2e^{-\sqrt{s/D}\,d}}{R_i d_{ij}\sqrt{s/D}}\,\sinh\left(\sqrt{s/D}\,R_i\right) + 2 a_0 \frac{e^{-\sqrt{s/D}\,R_i}}{\sqrt{s/D}\,R_i},
\end{equation}
thus
\begin{equation}
    a_0 = -\frac{\hm_j}{4\pi D d_{ij}}\,e^{-\sqrt{s/D}\left(d_{ij}-R_i\right)}\,\sinh\left(\sqrt{s/D}\,R_i\right).
\end{equation}
Introducing this value of $a_0$ together with $b_0 = 0$ into (\ref{eq:hC_general_solution}) and differentiating we get
\begin{equation}
    \int_0^\pi \left.\frac{\partial \hC}{\partial r}\right|_{r=R_i}\,\sin\theta\,\mathrm{d}\theta = -\frac{\hm_j}{2\pi D d_{ij} R_i} \exp\left(-\sqrt{\frac{s}{D}}\left(d_{ij}-R_i\right)\right).
\end{equation}
Transforming this expression back into the time domain and introducing it in the definition of the mass flux (\ref{eq:mass_flux}),
\begin{equation}
    \dot{m}_{ij} = -\frac{R_i^2 D}{2\sqrt{\pi}}\int_0^t \frac{1}{R_i}\left(1-\frac{R_i}{d_{ij}}\right)\frac{\dot{m}_j(t')}{\left[D(t-t')\right]^{3/2}}\,\exp\left(-\frac{(d_{ij}-R_i)^2}{4D(t-t')}\right)\,\mathrm{d}t'.
    \label{eq:mass_flux_time}
\end{equation}
To obtain the total mass flux into bubble $i$, this expression must be summed over all bubbles in the cloud (with the exception of $i$ itself) and then added to the mass flux provided by the Epstein--Plesset equation, $\dot{m}_{i0}$, to yield
\begin{equation}
\begin{split}
    \dot{m}_i & = 4\pi R_i^2 D \left(C_\infty - C_s\right)\left(\frac{1}{R_i} + \frac{1}{\sqrt{\pi D t}}\right) -\\ &\frac{R_i^2 D}{2\sqrt{\pi}}\sum_{\substack{j=1 \\ j\neq i}}^N\int_0^t\frac{1}{R_i}\left(1-\frac{R_i}{d_{ij}}\right)\frac{\dot{m}_j(t')}{\left[D(t-t')\right]^{3/2}}\,\exp\left(-\frac{(d_{ij}-R_i)^2}{4D(t-t')}\right)\,\mathrm{d}t'.
\end{split}
\label{eq:eqn_mdot_dim}
\end{equation}

To make the above expressions dimensionless we introduce the following variables (suppressing indices), $a = R/R_c$, $\Delta = d/R_c$, $\tau = D t/ R_c^2$ and $\dot{M} = \mathrm{d}M/\mathrm{d}\tau = \dot{m}/(\rho_g D R_c)$, where $R_c$ is a fixed characteristic bubble diameter, and $\rho_g$ is the gas density at the ambient pressure $P_s$.
Besides, the saturation level, which indicates the amount of CO$_2$ dissolved in the liquid, is defined by $\zeta = C_\infty / C_s$. Let us define also the Ostwald coefficient as $\Lambda = C_s/\rho_g$. With these definitions, equation (\ref{eq:eqn_mdot_dim}) results in
\begin{equation}
\begin{split}
    \dot{M}_i & = 4\pi a_i^2 \Lambda\left(\zeta-1\right)\left(\frac{1}{a_i} + \frac{1}{\sqrt{\pi\tau}}\right) -\\ &\frac{a_i^2}{2\sqrt{\pi}}\sum_{\substack{j=1 \\ j\neq i}}^N\int_0^\tau \frac{1}{a_i}\left(1-\frac{a_i}{\Delta_{ij}}\right) \frac{\dot{M}_j(\tau')}{\left[\tau-\tau'\right]^{3/2}}\,\exp\left(-\frac{(\Delta_{ij}-a_i)^2}{4(\tau-\tau')}\right)\,\mathrm{d}\tau'.
\end{split}
\label{eq:eqn_mdot}
\end{equation}
This mass flux must be completed with the mass conservation equation for the $i-$th bubble, namely
\begin{equation}
    \dot{M}_i = 4\pi a_i^2 \frac{\mathrm{d}a_i}{\mathrm{d}\tau}.
    \label{eq:eqn_massconservation}
\end{equation}

The changes in the volume of the bubble induce a velocity field which, sufficiently far away from its center, can be modelled as that of a volume point source. Thus, considering the whole cloud, each individual bubble moves as a result of the superposition of these flows. In the Stokes limit, the resulting bubble velocity can be computed using the solution provided by \citet{Michelin_etalPRF2018}
\begin{equation}
    {\bar{U}_i} =  \frac{1}{4\pi} \sum_{\substack{j\neq i}}^N \frac{\mathrm{\bar{e}_{ij}}}{\Delta_{ij}^2} \left( \dot{M}_j -\sum_{\substack{j\neq l}}^N \frac{a_j^3  \dot{M}_l}{\Delta_{jl}^3}\left(1-3(\mathrm{\bar{e}_{ji}} \cdot \mathrm{\bar{e}_{jl}})^2\right) \right) .
    \label{eq:velocities}
\end{equation}
In this equation, $\bar{e}_{ij}$ is the unit vector pointing from the center of the $i$-th to that of the $j$-th bubble. In this work, we are going to let bubbles translate with only the first term of this expression since we expect $\left(a_{j}/\Delta_{jl}\right)^3 \ll 1$.

\subsection{Numerical method}

The computation of the time evolution of the bubble cloud is divided into two stages. First, we integrate equations (\ref{eq:eqn_mdot}) and (\ref{eq:eqn_massconservation}) together assuming that the bubbles do not displace and then, in a second stage, we compute the motion of the bubbles using the velocity field given by (\ref{eq:velocities}).

The system of integro-differential equations (\ref{eq:eqn_mdot}-\ref{eq:eqn_massconservation}) is solved using an explicit Euler method for the temporal derivatives and a trapezoid method for the integral. Note that the kernel of the integral equation tends smoothly to zero at the integration limit $\tau' = \tau$, which makes the usage of more sophisticated integration methods unnecessary. We can distinguish two terms in equation (\ref{eq:eqn_mdot}): the first one is the mass flux provided by the Epstein-Plesset equation\cite{EpsteinPlessetJCP1950}, which corresponds to the evolution of an isolated bubble. The second term, the integral, models the interaction between each individual bubble and the rest of the cloud. For the calculation of this second term we will adopt the simplifying assumption that the radius of the bubble corresponds to the initial one (frozen bubble). Note that this hypothesis is also adopted in the derivation of the Epstein--Plesset equation\cite{EpsteinPlessetJCP1950}. The time step chosen for the simulations presented in this paper is $\Delta\tau = 10^{-3}$, which has been selected after verifying that taking a smaller time step does not affect the results.

Once the time evolution of the bubble radii is computed, the bubbles are displaced using the velocity field given by (\ref{eq:velocities}) using also, for consistency with the previous stage, an Euler method with the same time step.

The start-up of the numerical computation is done by integrating solely the Epstein--Plesset equation up to the time step used for the rest of the calculation, $\Delta\tau$. Note that this is possible, since at very short times the concentration boundary layer is confined to a thin shell around the bubble of thickness $\sim \sqrt{\tau}$, which is much smaller than the typical bubble-bubble distance. This makes it possible to neglect the integral term in this first step. This strategy is also used in the computation of the bubble velocities.

\section{Results and discussion}\label{sec:results_discussion}

We split this section into two parts. First, we explore the predictions of the theoretical model just described for conditions representative of our experiments. Then, we compare these predictions, in a qualitative way, with those features of the evolution of the bubble clouds that could be determined from the experiments.

\subsection{Results of the model}

In this section we show the predictions of the model for bubble clouds representative of those observed in the experiments. Since the number of dimensionless parameters of the problem is large, we restrict our results to initially spherical bubble clouds with a number of bubbles large enough to exhibit collective effects, namely $N=100$, and evolving in liquids with different saturation levels, $\zeta$. Furthermore, all the bubbles have the same initial size, $R_0 = 30$ ${\mu}$m. This radius is consistent with the estimated size of the bubble fragments that result in a similar cavitation experiment\cite{rodriguez2014}. 
Note that for bubbles of this size it is reasonable to neglect the effect of surface tension in their dynamics, as the Laplace pressure, $2\sigma/R_0$, with $\sigma \approx 0.7$ N/m the surface tension coefficient, only becomes of the order of the ambient pressure for bubbles of size $2\sigma/P_s \approx 1.4$ $\mu$m. In all the calculations we have adopted $\Lambda = 0.842$ and $D = 1.92\times 10^{-5}$ m$^2$/s, corresponding to the values of the system CO$_2$-water at 25$^\mathrm{o}$C\cite{PenasLopez_etalJFM2017}.
The initial positions of the bubbles are chosen randomly within a sphere, the cloud, of radius $R_{bc}$ such that the average bubble density is approximately uniform. To clearly identify the moment when bubbles start to interact with each other, it is convenient to seed the bubbles in the cloud with an average inter-bubble nearest neighbor distance, $d$, larger than, but still of the order of, one bubble diameter. Assuming that the volume of cloud available per bubble is $V_{\rm b} \sim \pi d^3 / 6$, setting $R_{bc} \approx 10 R_0$ yields $d \approx 4.3 R_0$, which fulfills the above criterion. In what follows, all length scales are made dimensionless with $R_0$ and time scales with $R_0^2/D$, using the same notation as in the previous section.

\begin{figure}
    \centering
    \includegraphics[width=\columnwidth]{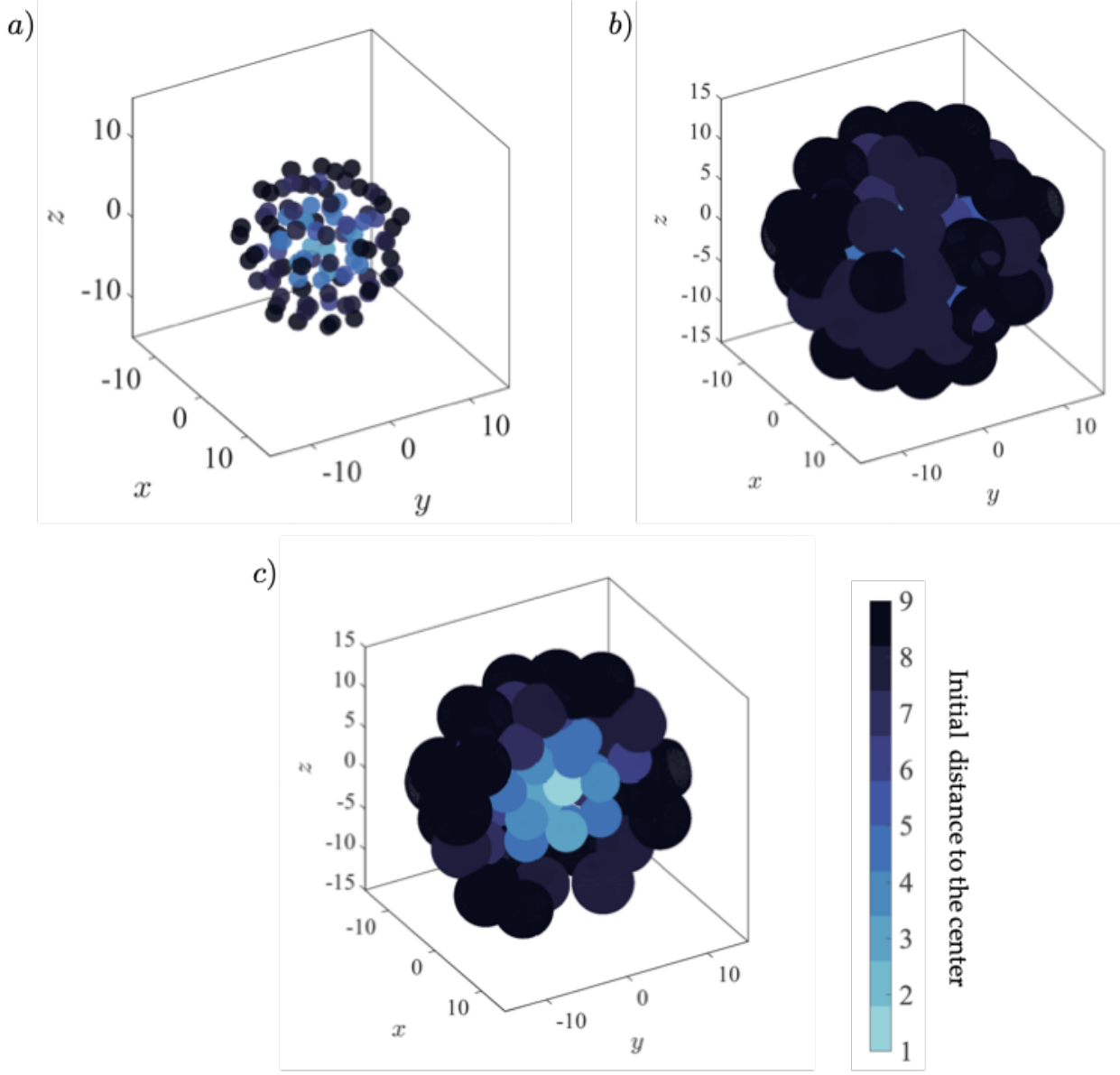}
   \caption{\label{fig:Initial_final_cloud}. Three-dimensional view of a bubble cloud generated with the model described in section 3, ($a$) at time $\tau=0$ and ($b$) at a later time, $\tau_{end}=3.2$, in dimensionless units. Panel ($c$) represents half the cloud, consisting of only those bubbles with centers obeying $x<0$, at the same time. The saturation of the liquid is $\zeta = 2$. The color represents the initial distance of each bubble to the center of cloud, growing from lighter to darker colors.}
\end{figure}

Figure \ref{fig:Initial_final_cloud} shows qualitatively the evolution of a modeled bubble cloud with a saturation $\zeta = 2$. The top left side is the initial state, $\tau=0$, whereas the top right one corresponds to the cloud at a later time $\tau = 3.2$. The color indicates the initial distance from the bubble to the center of the cloud, growing from light to dark hue. To illustrate this color coding, the bottom image represents a cut of the cloud at $\tau = 3.2$. To average results for bubbles located at a given initial distance from the center, the cloud radius in the initial state has been divided into $B=10$ layers equally spaced along the radius $R_{bc}$, with the color indicating the layer number. Using this averaging strategy we can show how bubbles grow at different depths inside the cloud.

\begin{figure}[h!]
    \includegraphics[width=\columnwidth]{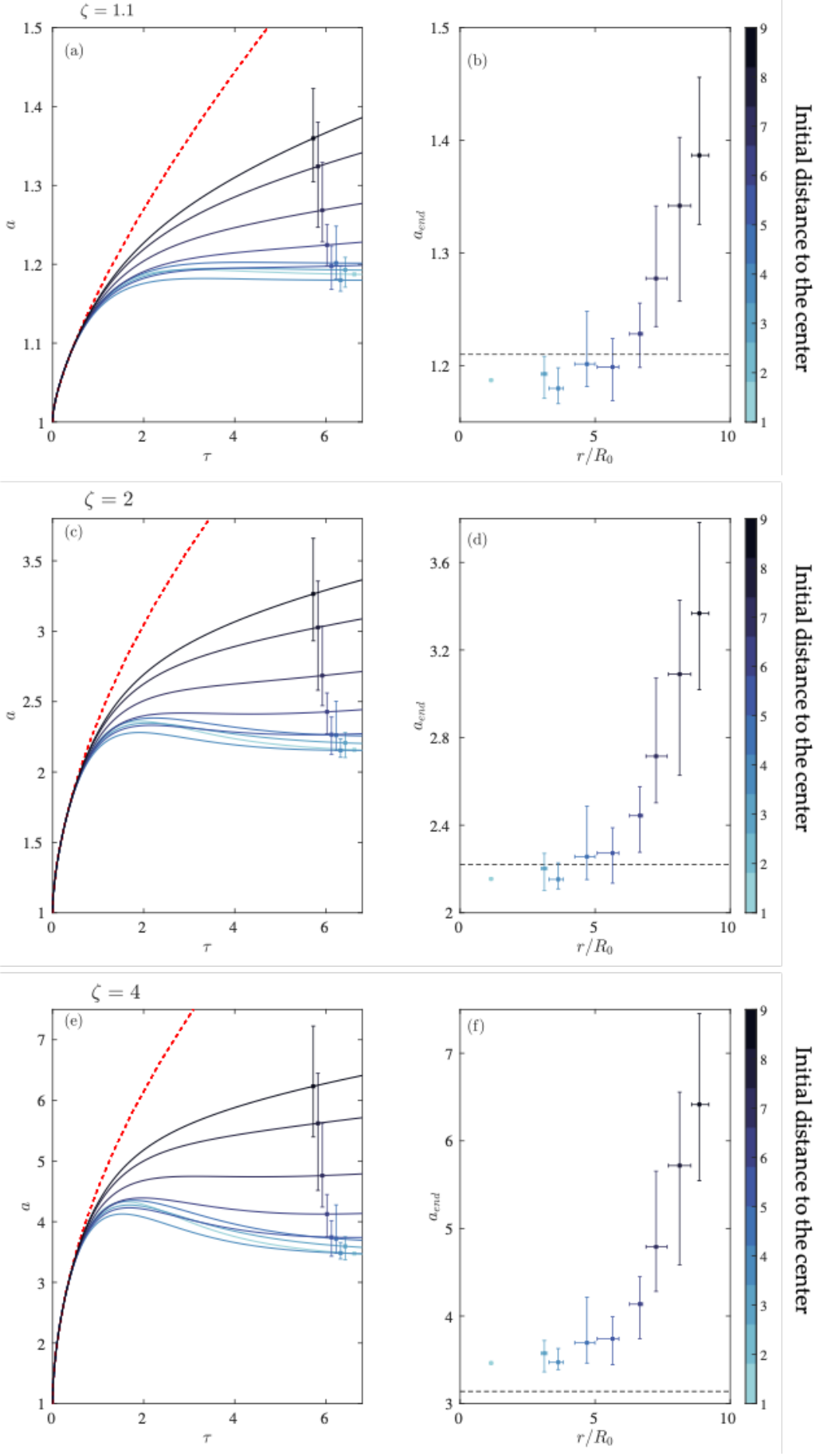}
    \caption{($a, c, e$) Dimensionless time evolution of the bubble radii for saturation levels $\zeta = 1.1, 2$ and $4$ respectively. The solid-line shows the mean radius for each cloud layer, with the color evolving from lighter to darker according to the distance of the layer to the cloud center. The error bars mark the maximum and minimum radius of the bubbles for each layer at that time instant. The red-dashed line represents the growth of an isolated bubble predicted by Epstein-Plesset equation \cite{EpsteinPlessetJCP1950}. ($b, d, f$) Final layer-averaged radius of the bubbles in the cloud as a functions of its distance to the center. Markers represent the mean final radius, whereas the vertical bars show the maximum and minimum value of the radius. The horizontal error bars correspond to the maximum and minimum final distance to the cloud center of the bubbles contained in each layer. The black horizontal dashed line is the value of $a_f$ predicted by equation (\ref{eq:af_estimation}).}
    \label{fig:radii_table}
\end{figure}

Panels ($a$, $c$, $e$) of figure~\ref{fig:radii_table} show the time evolution of the radii of the bubbles in three clouds with saturation levels $\zeta = 1.1, 2$ and 4, averaged over each radial layer. Correspondingly, panels ($b$, $d$, $f$) show the average bubble radius of each layer at the end of the simulation ($\tau \approx 6.5$). The duration of the simulation has been chosen to allow for the observation of the different growth regimes. At early stages all the bubbles grow freely, that is, as isolated bubbles. In this situation, their growth is well described by the Epstein--Plesset equation \citep{EpsteinPlessetJCP1950}, corresponding to the red dashed line in the figure. Using our notation, this equation reads

\begin{equation}
    \frac{\mathrm{d}a}{\mathrm{d}\tau} = \Lambda\left(\zeta-1\right)\left(\frac{1}{a} + \frac{1}{\sqrt{\pi\tau}}\right).
    \label{eq:epstein_plesset_dimensionless}
\end{equation}
The departure from the growth predicted by this equation is a consequence of the fact that the thickness of the diffusive boundary layer around the bubble grows as $\delta \sim \sqrt{\tau}$. Therefore, until times of order $\tau \sim \left(\Delta/2 - 1\right)^2$ the boundary layers of the different bubbles do not interact and bubbles do not feel each other. However, from that time onwards, the diffusive boundary layers overlap and bubbles compete for the available CO$_2$. As a consequence, their growth rates diminish quickly, even becoming zero or slightly negative for bubbles in the innermost layers, where the access to the CO$_2$ present in the bulk outside the cloud is limited. Contrarily, the growth of the bubbles in the husk of the cloud continues, albeit at a much smaller pace. Note that the freezing of the growth of the innermost bubbles, which is also clearly visible in figure~\ref{fig:Initial_final_cloud}c, is consistent with the experimental observations (see figure \ref{fig:minibubbles_exp} and the associated discussion). Regarding the slight reduction of the bubble sizes for bubbles deep inside the cloud for $\tau \gtrsim 2$, observed in figure~\ref{fig:radii_table}, although it seems reminiscent of Ostwald-rippening, where surface tension causes large bubbles to grow at the expense of small ones\citep{Schmelzer1987}, surface tension is not considered here. Consequently, this bubble size reduction cannot be attributed to this effect. In fact, in the absence of surface tension, the gas concentration at the bubble surface could never be smaller than the saturation one, $C_s$, which means that bubbles should never shrink. The cause of this anomalous effect is the fact that we model nearby bubbles as point sinks. For a point sink, imposing a given mass flux means that the concentration is minus infinity at the sink. So, if the sink is very close to the bubble, it may occur that locally the concentration becomes smaller than the saturation one, which is a non-physical effect. Naturally, this would not occur were the sink replaced by a physical bubble, as in its surroundings the concentration verifies $C \ge C_s$ at all times. Nonetheless, the bubble size reduction created by this simplification is at most about 15\% in the case $\zeta = 4$ and smaller in the others.

The final radius of bubbles belonging to the innermost layers can be estimated using simple mass conservation arguments. Since these bubbles lose access to the CO$_2$ from the bulk liquid, they almost stop growing when they have absorbed all the excess CO$_2$ dissolved in the liquid volume of cloud available per bubble, which we can regard as that of a sphere of diameter $d$. Thus, the final dimensional mass, $4\pi/3 \, R_f^3 \rho_g$, must the the sum of the initial one, $4\pi/3 \, R_0^3 \rho_g$, plus the excess mass of CO$_2$ dissolved in the liquid volume with respect to the saturation one, namely $(C_\infty - C_s)\pi/6 \, d^3$. In dimensionless terms this yields for the final radius
\begin{equation}
    a_f = \left(1 + \Lambda\left(\zeta-1\right)\Delta^3/8\right)^{1/3}.
    \label{eq:af_estimation}
\end{equation}
For the saturation levels corresponding to the simulations in figure~\ref{fig:radii_table}, namely $\zeta = 1.1, 2$ and 4, the values of $a_f = 1.21, 2.22$ and 3.14 respectively. We must point out here that when we refer to final radius or bubble sizes, we mean at the end of the simulation. Note that, even at the center of the bubble cloud, still bubble radii keep changing with time at long times. However, they do so at a much smaller pace than that predicted by the Epstein--Plesset equation, thus we find it reasonable to regard it as quasi-frozen.

As can be observed in figures~\ref{fig:radii_table}$b$, $d$ and $f$, the agreement with the numerical results is fairly reasonable, given the assumptions adopted in this estimation. In this figure, markers denote the layer-averaged final radius, with the vertical error bars indicating the minimum and maximum bubble size for each layer. In an analogous way, the horizontal position of each marker represent the mean radial position of the bubbles in the layer at the end of the simulation, which are all contained in the segment marked by the horizontal error bars.

\begin{figure}[h]
\centering
\includegraphics[width=0.7\columnwidth]{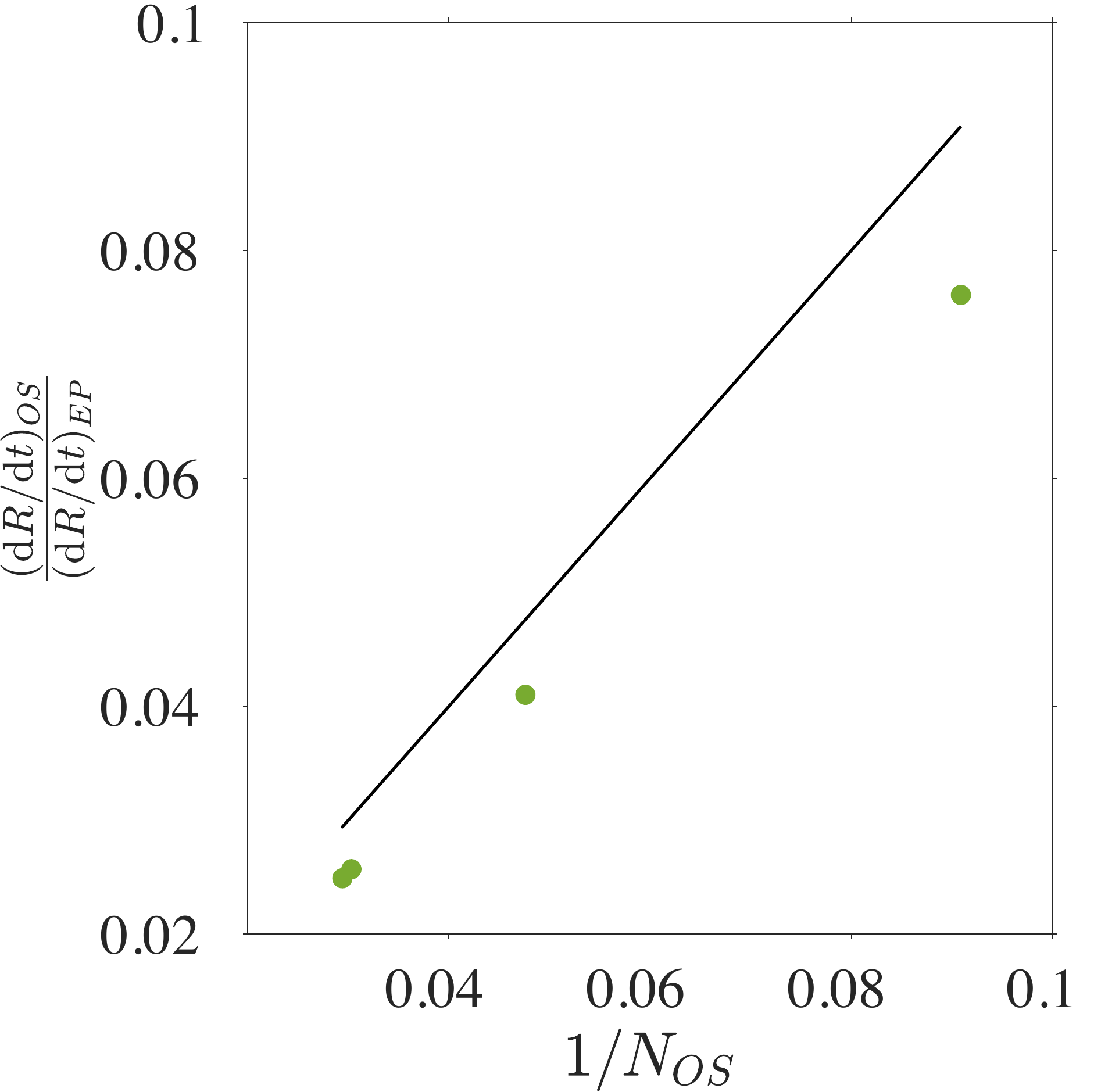}
\caption{Rate of change of the radius of the bubbles in the outermost layer divided by that predicted by the Epstein--Plesset equation, as a function of the inverse of the number of bubbles in this layer. The solid line is $1/N_{os}$.}
\label{fig:Nos_Rep}
\end{figure}

Once bubbles deep into the cloud reach this maximum size, it seems reasonable to assume that only those in the outer shell continue growing. Because there is almost no CO$_2$ inside the cloud, all the mass flux comes across the outer boundary of the cloud. Moreover, this mass flux must be shared between all the bubbles in the outer shell. In a first approximation, we can model this effect by assuming that each bubble does no longer intake gas from a solid angle $4\pi$ (that is, from all directions) but from a solid angle $4\pi / N_{os}$, where $N_{os}$ is the number of bubbles in the outer shell. This means that the Epstein--Plesset equation can still be used to model bubble growth in the husk, but dividing its right hand side by the number of bubbles in the outer shell. To check this hypothesis, we plot in figure~\ref{fig:Nos_Rep} the time derivative of the radius of the bubbles in the outermost layer at the end of the simulation divided by the one predicted by the Epstein--Plesset equation as a function of the inverse of the number of bubbles in this layer. These results correspond to the same saturation level, $\zeta = 2$, but to four different realizations. Because in each realization the bubbles are distributed randomly, the number of them in the outermost layer varies. The average value of this number can be estimated by assuming that the bubble density in the cloud is constant and that the outermost layer has a thickness of about $d/2$. Using simple geometrical analysis this yields $N_{os} \approx 3Nd/2R_{bc}$.

\begin{figure}[h]
    \centering
    \includegraphics[width=0.7\columnwidth]{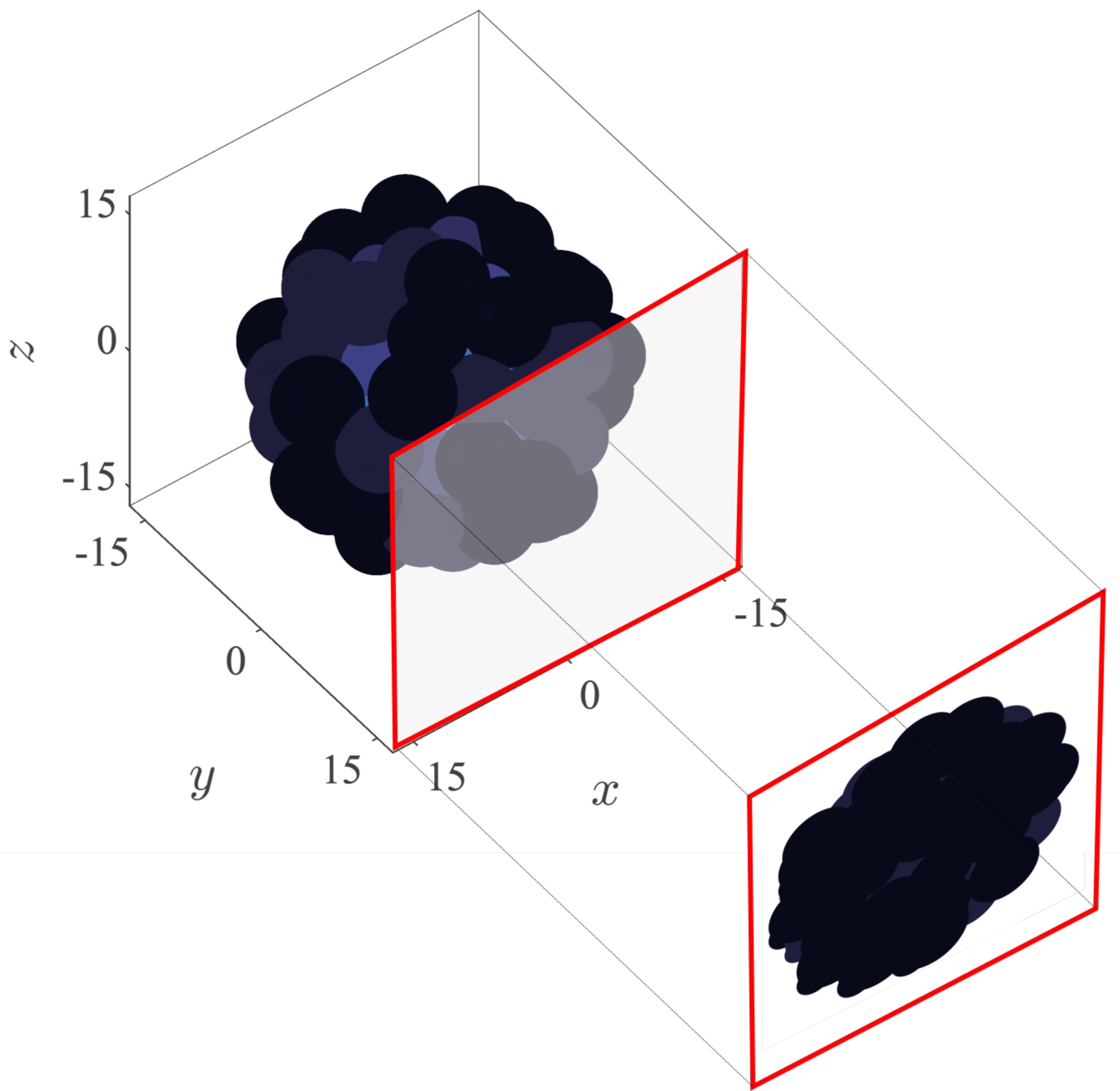}
    \caption{Schematic of the definition of the projected area.}
    \label{fig:sketch_projected_area}
\end{figure}

We conclude this section with a comment about the effect of the saturation level on the growth rate of the cloud. Although in the experiments the radius of the individual bubbles cannot be determined experimentally, it is possible to measure the time evolution of the projected area of the whole cloud by applying image processing to each frame of the high-speed movies. To compare these results against the simulations, we need to compute the time evolution of the projected area of the simulated cloud. To that end, we create synthetic images projecting the three-dimensional bubble cloud onto a plane, as sketched in figure~\ref{fig:sketch_projected_area} and done in figure~\ref{fig:Initial_final_cloud}. Using this technique, we are able to take into account the effect of bubbles overlapping on the projected cloud area. Note that, since the cloud is spherical and the bubbles are distributed randomly, the choice of the projection plane does not have an effect on the results.

{}

\begin{figure}[h]
    \centering
    \includegraphics[width=0.7\columnwidth]{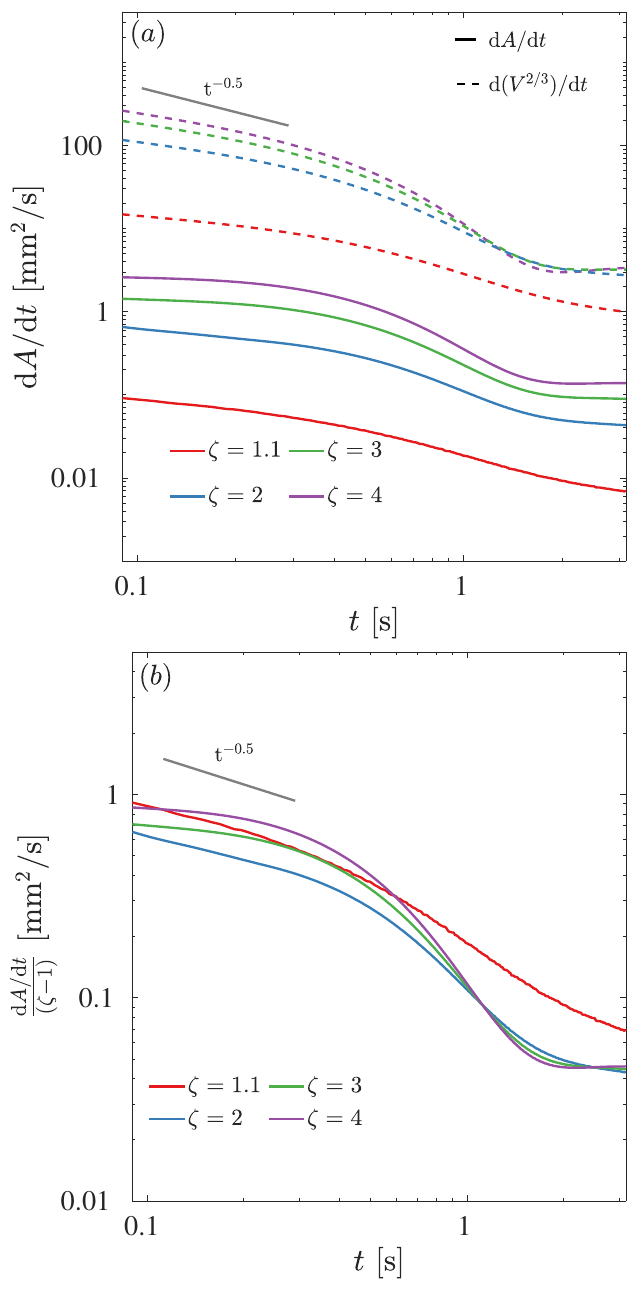}
    \caption{($a$) Time derivative of the measured projected area and the measured 2/3 root of the volume for numerically created bubble clouds with identical initial configurations but different saturation levels, $\zeta$. ($b$) Time derivative of the projected area divided by $(\zeta-1)$. Note that, to facilitate the comparison with the experiments, we show here the numerical results in dimensional form, using $R_0 = 30 \mu$m and $R_0^2/D \approx 0.47$ s as length and time scales respectively.}
    \label{fig:dAdt_sim_different_zetas}
\end{figure}

The physics behind the growth of the bubble cloud becomes more clear if we focus on the time derivative of the projected area rather than on the area itself, as we explain in the next subsection. Figure~\ref{fig:dAdt_sim_different_zetas}$a$ shows how, for $\zeta \ge 2$, the evolution of both the projected area and the total gas volume is qualitatively independent of this value, just affecting the absolute value of the growth rates.

Indeed, examination of the Epstein--Plesset equation (\ref{eq:epstein_plesset_dimensionless}) suggests that, provided the radius of the bubble is sufficiently larger than the initial one, we can assume $a \sim \sqrt{\tau}$. In this case, the growth rate of $a^2$, which is roughly proportional to the projected area, scales with $\left(\zeta-1\right)$. To check this scaling, we plot in Figure~\ref{fig:dAdt_sim_different_zetas}$b$ the time derivative of the dimensional projected area divided by the saturation level, $\zeta - 1$. As predicted, the curves collapse fairly well at intermediate and long times. At short times, because the bubble radii are still close to the initial ones, the collapse is not so good. Furthermore, the curve corresponding to the smallest saturation level, $\zeta = 1.1$, does not exhibit such a good collapse at any time. This is due to the influence of the initial bubble radius, which lasts longer due to the smaller growth rate.

The model predicts the existence of three different stages in the evolution of the cloud which, despite their noise and variability, are also observed in the experiments. At short times, the area grows relatively fast, at a nearly constant rate. Later on, the rate of change of the area decays slightly faster than the square root of the time. Finally, at long times, the area grows again at a constant rate, albeit at a much slower pace. 

The last two regimes can be easily explained with a simple heuristic model. Let us assume that the volume of the cloud is the sum of that of the $N$ bubbles plus their associated water volume (a sphere of radius $d/2$ for each bubble). Since the water volume in the cloud does not change, the radius of such a cloud would then be
\begin{equation}
    a_{bc} = \left(N\left(\frac{d}{2}\right)^3 + \sum_i a_i^3\right)^{1/3} = N^{1/3}\left(\left(d/2\right)^3 + a^3\right)^{1/3}.
\end{equation}
where, for simplicity, we take all bubbles to have the same size.
The projected area, $A_{bc} = \pi a_{bc}^2$, has a rate of change
\begin{equation}
    \frac{\mathrm{d}A_{bc}}{\mathrm{d}\tau} = 2\pi N^{2/3} \frac{a^2 \dot{a}}{\left(\left(d/2\right)^3 + a^3\right)^{1/3}}.
    \label{eq:theoretical_rate_change_A}
\end{equation}
At short times, the radius of the bubble does not differ much from its initial one, $a \approx 1$, whereas its time derivative goes as $\dot{a} \sim \tau^{-1/2}$, as predicted by the Epstein--Plesset equation. Introducing this into Equation (\ref{eq:theoretical_rate_change_A}), we get $\mathrm{d}A_{bc}/\mathrm{d}\tau \sim \tau^{-1/2}$. This is in fairly good agreement with the behavior observed at intermediate times in both experiments and theory. Actually, the rate of change of the area seems to decay slightly faster than this prediction, which is consistent with the fact that bubbles deep inside the cloud stop growing once they deplete their surroundings from CO$_2$. In other words, the effective value of $N$ in Equation (\ref{eq:theoretical_rate_change_A}) also decays with time, until it only refers to those bubbles located in an outer shell of thickness of order $\Delta$.

At long times, the radius of the bubbles in the outer shell still grows as $a \sim \tau^{1/2}$, albeit with a prefactor smaller than that predicted by Epstein--Plesset, as discussed above. Moreover, $a \gg d/2$. In this limit, Equation (\ref{eq:theoretical_rate_change_A}) yields $\mathrm{d}A_{bc}/\mathrm{d}\tau \sim a \dot{a}$, which is constant. As we will see in the next subsection, this is indeed what is observed to occur in the simulations and is qualitatively consistent with the experiments as well, despite the noise.

At short times, the model predicts a fast growth rate, not only in absolute value, which is to be expected as a consequence of bubbles growing without competing between them, but also in terms of how slowly it decays. Indeed, although Equation (\ref{eq:theoretical_rate_change_A}) suggests that $\mathrm{d}A_{bc}/\mathrm{d}\tau \sim \tau^{-1/2}$ at short times as well, we observe that $\mathrm{d}A_{bc}/\mathrm{d}\tau$ is rather constant instead. This is a consequence of the fact that the void fraction inside the cloud grows fast at short times. Thus, the growth of the projected area is the addition of two effects: its boundaries grow, as predicted by Equation (\ref{eq:theoretical_rate_change_A}), and the interior of the projected cloud image becomes more opaque as more bubbles overlap on the projection.
Naturally, if we compute the total gas volume inside the cloud, $V$, we observe a much closer agreement with $\mathrm{d}V^{2/3}/\mathrm{d}t \sim \tau^{-1/2}$, since the fact that bubbles overlap in the image does not affect the evolution of the volume. This can be observed in figure~\ref{fig:dAdt_sim_different_zetas}, where the evolution of the time derivative of $V^{2/3}$ is compared with that of the projected area of the cloud $A$ for simulations with identical initial bubble distributions but different saturation levels.

\subsection{Comparison with experiments}

In this section we analyze the evolution of the size of the bubble clouds observed in our microgravity experiments, comparing it with that predicted by the model. Several key features of the bubble cloud, such as the number of bubbles or their initial size and spatial distributions cannot be determined experimentally. For this reason, the comparison will be predominantly of qualitative nature and restricted to the area of the bubble cloud projected on an image, which is the magnitude that can be measured experimentally. Nonetheless, despite this limitation, we will show that the different regimes and growth rate behaviors predicted by the model are consistent with those observed experimentally.

An important aspect is to determine the saturation level of CO$_2$ of the water in the experiments. Some isolated bubbles appear in almost all the experiments, and we use them to determine the saturation level, $\zeta$, as done in \cite{VegaMartinez_etalMGST2017}. To this end, we track the evolution of their radii by imaging processing. Then, we estimate $\zeta$ by fitting that evolution to that predicted by the Epstein--Plesset equation \cite{EpsteinPlessetJCP1950}, obtaining values between 3.5 and 4.3, depending on the experiment.

\begin{figure}[h]
    \centering
    \includegraphics[width =0.5\columnwidth]{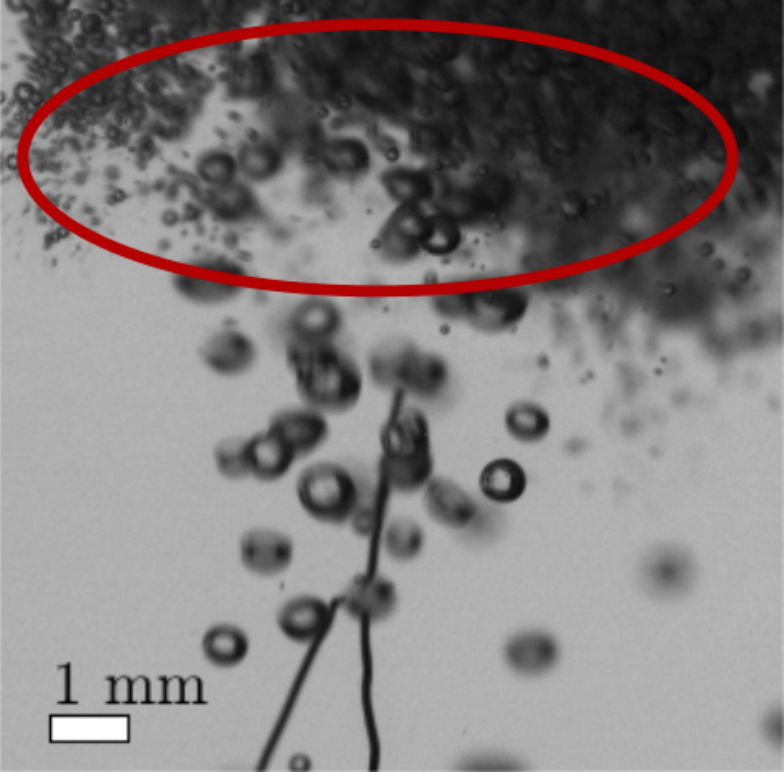}
    \caption{Snapshot of a high-speed movie showing the cloud during the deceleration of the capsule ($\zeta \approx 4$). As the capsule comes to a stop, the large deceleration makes the bubble cloud rise fast. Because large bubbles rise faster, this allows us to see those that were inside the cloud (within the red ellipse region). See also movie in the supplemental material.}
    \label{fig:minibubbles_exp}
\end{figure}

As stated above, the high bubble density of the cloud precludes the observation of its interior. Consequently, it is not possible to determine experimentally the number of bubbles nor their size distribution, at least not accurately. However, as soon as the capsule brakes upon reaching the bottom of the tower, the sudden appearance of a large apparent gravity allows us to instantly observe those bubbles that were at the center of the cloud (figure \ref{fig:minibubbles_exp}). This observation reveals that bubbles deep inside the cloud are much smaller than those near the outer edge, as predicted by the model (figure \ref{fig:radii_table}).

\begin{figure}[h]
    \centering
    \includegraphics[width=0.8\columnwidth]{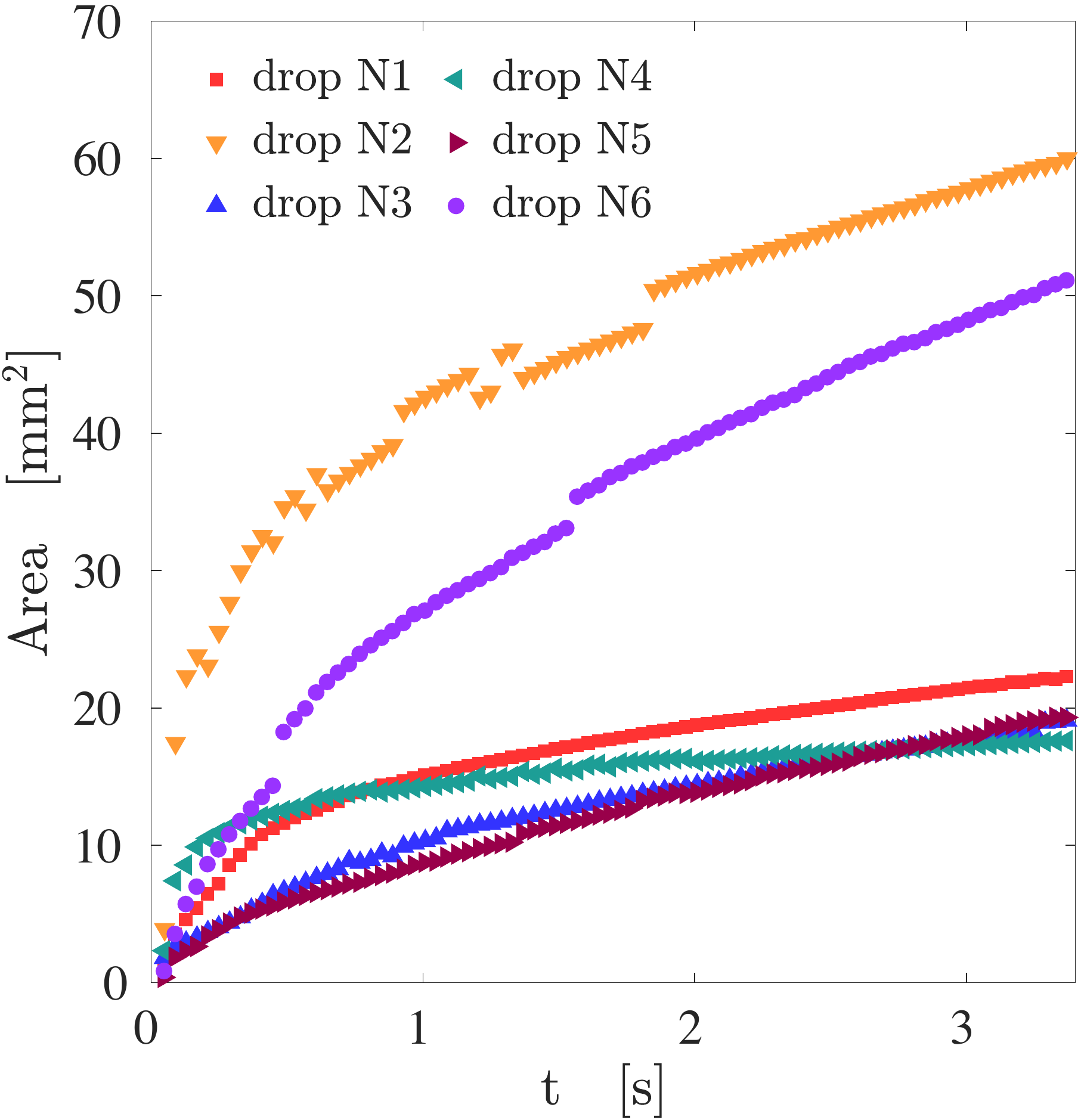}
    \caption{Evolution of the measured projected area of six bubble clouds. The level of saturation of the water in these experiments varies between 3.5 -- 4.3. }
    \label{fig:Areas_drops}
\end{figure}

Figure~\ref{fig:Areas_drops} shows the time evolution of the projected area for the six drops. In this and the following plot, all the magnitudes are plot with dimensions, to give an idea of the orders of magnitude of the spatial and temporal scales. The jumps observed in some of the curves are due to the merging or separation of individual bubbles from the main cloud.
Besides the fact that all the curves grow monotonically, it is not obvious to extract any trend, specially not the existence of different regimes. In part, this is a consequence of the large dispersion in the growth rates of the projected area, which arises from the variability in the initial conditions of the bubble cloud caused by the collapse of the spark-induced cavitation bubble. But also, even if all the clouds were generated identical, the projected area at a given instant is an integral of all the previous growth rates. Thus, this magnitude is not suitable to identify different stages in the cloud growth rate. For this reason, we focus our analysis on the time derivative of the projected area, which is an instantaneous magnitude, i.e. not depending on the bubble cloud history or initial conditions.

\begin{figure}
    \centering
    \includegraphics[width=0.8\columnwidth]{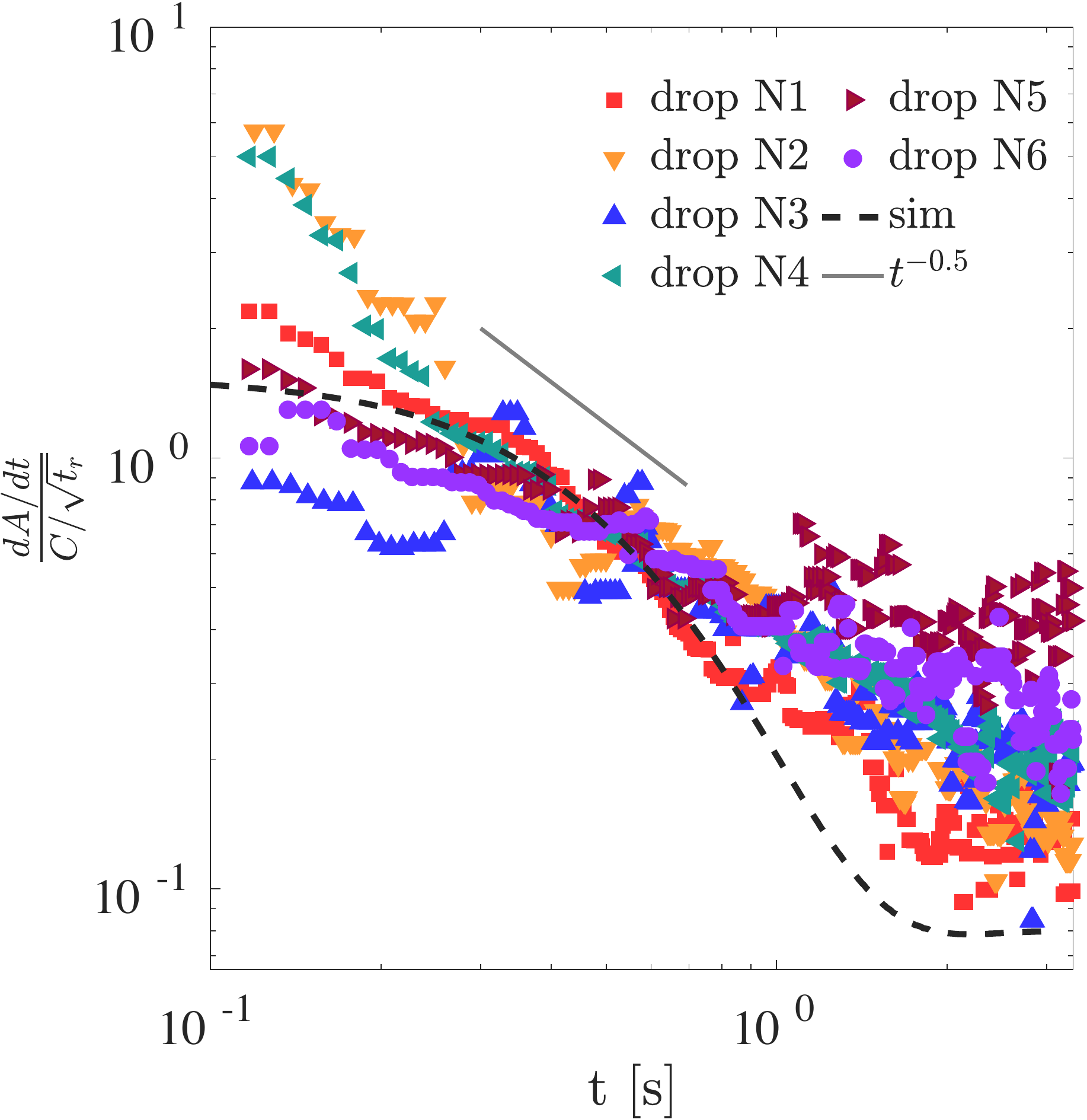}
    \caption{Rate of change of the projected area for the six drops, compared with that predicted by a simulation with $\zeta = 4$. The curves have been scaled with their corresponding value at the center of the time range where they follow the scaling law $\mathrm{d}A/\mathrm{d}t = C t^{-1/2}$, that is $t_r = 0.4$ s. Here, $C$ is the fitting constant of each experiment.}
    \label{fig:dAdt}
\end{figure}

Figure~\ref{fig:dAdt} shows in logarithmic scale the rate of change of the area of the cloud for the six drops carried out in the campaign (see appendix for details on how the time derivative of the area is calculated). The curves in the figure have been scaled with their value at the center of the time range where they follow the law $\mathrm{d}A/\mathrm{d}t \sim t^{-1/2}$ to be able to compare the evolution of clouds with very different initial sizes and thus growth rates. We plot also the rate of change of the area computed from a numerical simulation with $\zeta = 4$.

In the experiments, the behavior at short times exhibits a larger variability than that predicted by the model. In some experiments we observe the trend $\mathrm{d}A/\mathrm{d}t \sim t^{-1/2}$ from the first instants, whereas in others the projected area shows an initially constant growth rate. We attribute this anomalous behavior to the fact that, at such short times, some of the clouds still keep some residual velocity as a result of the implosion of the cavitation bubble which gave them birth. As this initial velocity dissipates, the projected area transitions to the regime where it decays as the inverse of the square root of the time. This residual velocity can be observed in the high-speed movies (See movies in the supplemental material). Finally, despite the experimental noise, it is possible to infer that the growth rates stabilizes and decays slower than $t^{-1/2}$, consistently with the model predictions.

\section{Conclusions}

We report here experiments where dense bubble clouds grow by diffusion in a supersaturated liquid under microgravity conditions. The bubble clouds are created upon the implosion of a cavitation bubble generated by a spark. Although this way of producing the clouds introduces some variability in the experiments, the evolution of the six bubble clouds studied is qualitatively identical, which suggests that the diffusive growth mechanism that we study is fairly independent of the particular size or shape of the cloud.
To achieve microgravity conditions, the experiments have been carried out in the drop tower facility of the German Center of Applied Space Technology and Microgravity (ZARM). The absence of gravity allows us to observe the purely diffusive growth dynamics of the bubble for several seconds, more than an order of magnitude longer than what could be observed under normal gravity\cite{rodriguez2014}, where bubble buoyancy would quickly become dominant. Moreover, we avoid other effects such as natural convection \cite{Enriquez_etalJFM2014, Moreno_soto_etalJFM2019} inside the liquid phase. 

Inspired by the experimental observations, we have developed a mathematical model where each bubble grows competing for the available CO$_2$ with the others in the cloud, which are treated as point mass sinks. Although, strictly speaking, the model is only valid in the limit of a dilute bubble cloud, it predicts reasonably well the qualitative evolution observed in experiments. The reasonably good agreement of the model with the observations has allowed us to explain, using simple heuristic arguments, the experimental observations. Moreover, this model has been used to investigate aspects that could not be determined experimentally, namely the final distribution of bubble sizes inside the cloud and the effect of the saturation level.

\section*{Acknowlegdements}
\textit{This work was supported by the Netherlands Center for Multiscale Catalytic Energy Conversion (MCEC), an NWO Gravitation programme funded by the Ministtry of Education, Culture and Science of the government of the Netherlands and the Spanish FEDER/Ministry of Science, Innovation and Universities--Agencia Estatal de Investigaci\'on though grants DPI2017-88201-C3-3-R and DPI2018-102829-REDT, partly funded through European Funds. We also acknowledge the support of the European Space Agency (ESA) for providing access to the drop tower through grants HSO/US/2015-29/AO and HRE/RS-PS/2018-7/AO. We thank the team from the ZARM Drop Tower Operation and Service Company (ZarM FAB mbH) for their valuable technical support during the experimental campaigns, and the technical staff of the Dept. of Thermal and Fluids Engineering of Universidad Carlos III de Madrid for their assistance in developing the experimental set up. We are indebted to Dr. L. Champougny for her valuable comments about the manuscript.}





\bibliography{lld2layers} 
\bibliographystyle{rsc} 

\newpage
\clearpage

\begin{appendices}

\begin{figure}
\begin{minipage}{\textwidth}
\section{Experimental data from all drops}

We present the post-processing analysis carried out for the high-speed movies obtained for the six drops. Regarding the image analysis, the projected area of the cloud have been tracked using custom-made image processing software implemented in Matlab. We started tracking this area from 2 or 3 frames after the spark generates the bubble cloud untill the end of the experiment. Thus, the time span during which the area is measured runs from 0.05 to 3.2 s. 

 As we said in section \ref{sec:results_discussion}, it is easier to observe the physics behind these experiments if we represent the time derivative of the projected area. To calculate the time derivative of the cloud size we use a central finite difference with a 4$^{th}$ order of accuracy with a uniform grid spacing, $\Delta$t = 0.001s, despite of our time experimental resolution is 5e-4 s, to avoid the numerical noise. Then, we screen the data using a ten-point centered moving mean filter in order to see the growth trends. The figure below shows the evolution of the projected area (first column) and the filtered time derivative of the two perpendicular views for the six drops.

\includegraphics[width=\textwidth]{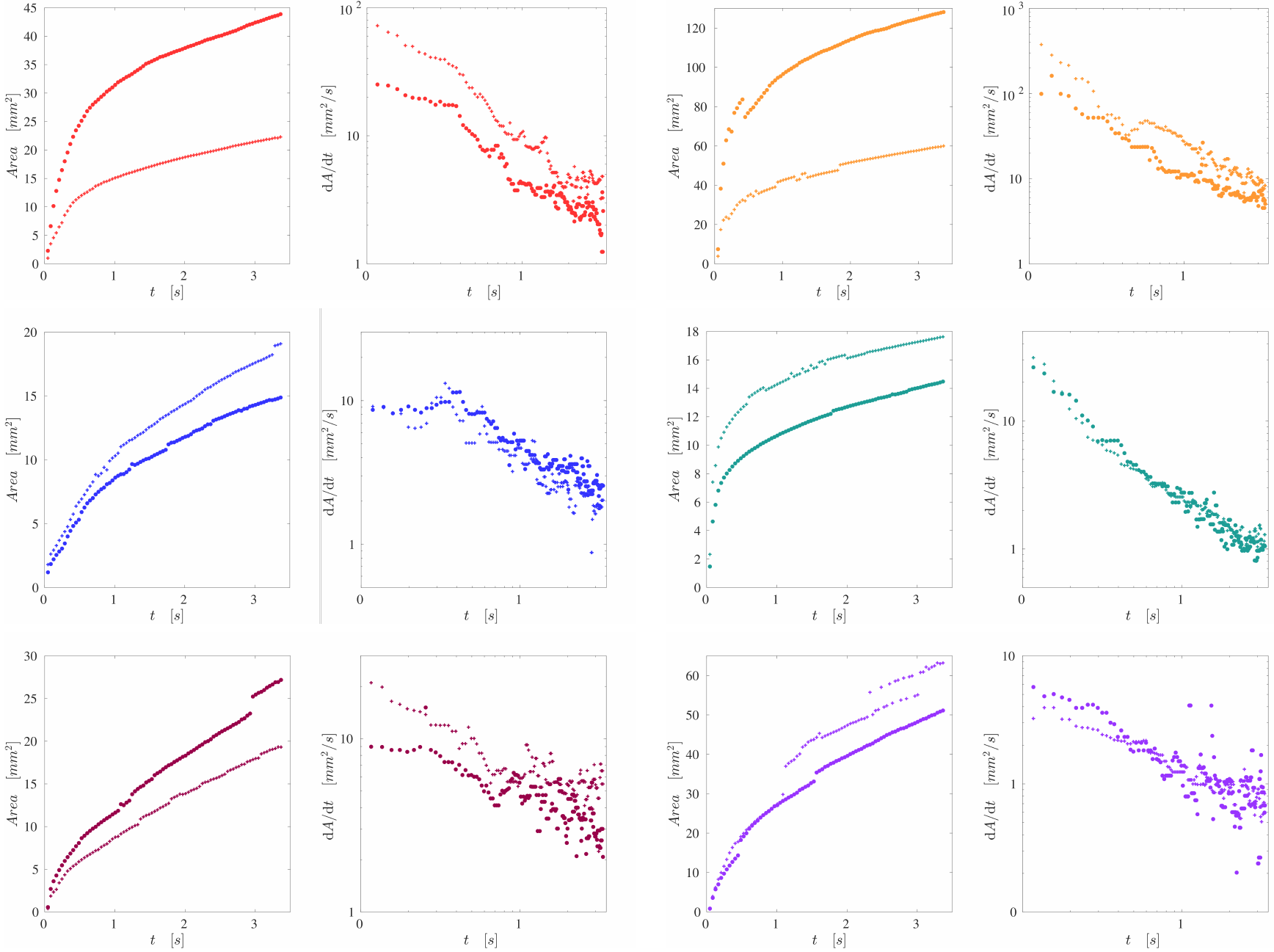}
 \vspace{2mm}

\end{minipage}
  \caption{Projected area and their time derivatives obtained from the two camera views in the six experiments carried out during the drop tower campaign.}
\end{figure}

\end{appendices}

\end{document}